\documentclass[a4paper,11pt]{article}

\usepackage{amssymb}
\usepackage{amsmath}
\usepackage{graphicx}
\usepackage{comment}

\begin{document}

\huge

\begin{center}
Self-consistent modelling of hot plasmas within non-extensive Tsallis' thermostatistics
\end{center}

\vspace{1cm}

\large

\begin{center}
Jean-Christophe Pain\footnote{jean-christophe.pain@cea.fr (corresponding author)}, Denis Teychenn\'e and Franck Gilleron
\end{center}

\normalsize

\begin{center}
CEA, DAM, DIF, F-91297 Arpajon, France
\end{center}

\begin{abstract}
A study of the effects of non-extensivity on the modelling of atomic physics in hot dense plasmas is proposed within Tsallis' statistics. The electronic structure of the plasma is calculated through an average-atom model based on the minimization of the non-extensive free energy.
\end{abstract}

\section{Introduction}
\label{intro}
For astrophysical applications, as well as for modelling laser-produced plasmas, equation-of-state data are required over a wide range of physical conditions. When the ions are strongly coupled (\emph{i.e.} when Coulomb interaction between ions becomes larger than the thermal kinetic energy), and the electrons degenerate ($k_BT \lesssim \epsilon_F$, where $\epsilon_F$ is the Fermi energy, $T$ the temperature and $k_B$ the Boltzmann constant), the plasmas combine features of hot matter, such as ionization and fluid behaviour, and characteristics of cold matter, such as electron degeneracy. Some electrons are bound to the nucleus, and others are free (or delocalized). The delocalized states are also sometimes called ``scattering states''. The electrons can be treated with various degrees of complexity. The idea of the average-atom model is to consider only the mean configuration of the atom. In this picture, the atom is viewed as confined to the Wigner-Seitz (WS) sphere, which is immersed in a homogeneous jellium of delocalized electrons neutralized by a continuous background. In the approach proposed by Rozsnyai \cite{ROZSNYAI72}, the bound electrons are treated quantum-mechanically, and the delocalized electrons as a Thomas-Fermi fluid \cite{FEYNMAN49}. This simple but efficient model has been used in numerous approaches of plasma structural and radiative properties. The first purely quantum approach to an average-atom model is due to Liberman \cite{LIBERMAN79}. In this model, the WS sphere appears explicitely as a cavity into which non-central ions can not enter. The delocalized electrons are treated quantum-mechanically, which results in Friedel-type oscillations of the self-consistent electron density and a potential extending beyond the WS radius. Recently, a new version of that approach, including efficient numerical methods, was developed by Wilson {\emph{et al.} \cite{WILSON06}. A great progress in the field was brought by R. Piron and T. Blenski \cite{PIRON11}, who developed a numerical code which provides the average-atom structure and the mean ionization self-consistently from variational equations. This work enables one to clarify the thermodynamic consistency issues in the existing average-atom models. For a sake of simplicity, in the present work we consider an average-atom model similar to the one proposed by Rozsnyai, in which bound electrons are treated quantum-mechanically (by Schr\"odinger equation with relativistic Pauli corrections \cite{BLENSKI95}), and delocalized electrons semi-classically (Thomas-Fermi model) \cite{PAIN07a}.

In standard thermodynamics of systems in local thermodynamic equilibrium (LTE), quantities such as energy and entropy are extensive, which means that they are proportional to the size of the system. The lack of adequacy of Boltzmann-Gibbs (BG) entropy is related to the breakdown of the extensivity. More precisely, BG statistics fails for instance when a system includes long-range interactions \cite{PLASTINO93,NOBRE95}, long-time memory effects (non-Markovian process) and/or evolves in a (multi)-fractal space. Tsallis' form has been widely used in various fields of physics and is considered as a possible framework to deal with non-extensive settings. Lima \emph{et al.} \cite{LIMA01,SILVA06} have studied the kinetic foundations of Tsallis' statistics through a modified Boltzmann transport equation satisfying a modified $H$theorem. 

When a physical system is made of a large number of identical sub-systems, the thermodynamic identity 

\begin{equation}\label{eq1}
\langle E\rangle-TS-\mu \langle N\rangle=-PV
\end{equation}

is verified, quantities $\langle E\rangle$ (average energy), $\langle N\rangle$ (average number of particles) and $S$ (entropy) are extensive and the grand potential provides a simple way to calculate the thermodynamic quantities. On the other hand, in the case of average-atom models, equation (\ref{eq1}) is \emph{a priori} not necessarily verified\footnote{It is also the case for the Virial theorem:

\begin{equation}\label{eqv}
3PV=2E_k+E_p
\end{equation}

which requires extensive quantities, and therefore a system made of a large number of identical sub-systems. In Eq. (\ref{eqv}), $P$ is the pressure, $V$ the volume, $E_k$ the kinetic energy and $E_p$ the potential energy.} since only one nucleus is taken into account and $\langle N\rangle$ must be equal to the atomic number $Z$. In the framework of the average-atom model, the system can not be understood as a statistical sum of many identical subsystems. 

In the present work, we propose to investigate how non-extensivity affects atomic-structure calculations in hot and dense plasmas within Tsallis' statistics. To our knowledge, it is the first time that non-extensivity and quantum shell-structure effects are taken into account simultaneously in a self-consistent procedure. In section \ref{sec2}, Tsallis' entropy is introduced \cite{TSALLIS88}. In section \ref{sec3}, the self-consistent calculation of the electronic structure (energies, occupancies and wavefunctions of the orbitals, etc.) is presented. The impact of non-extensivity on ionic distributions is illustrated in section \ref{sec4}. Section \ref{sec5} is the conclusion.

\begin{figure}
\vspace{0.5cm}
\includegraphics[width=12cm]{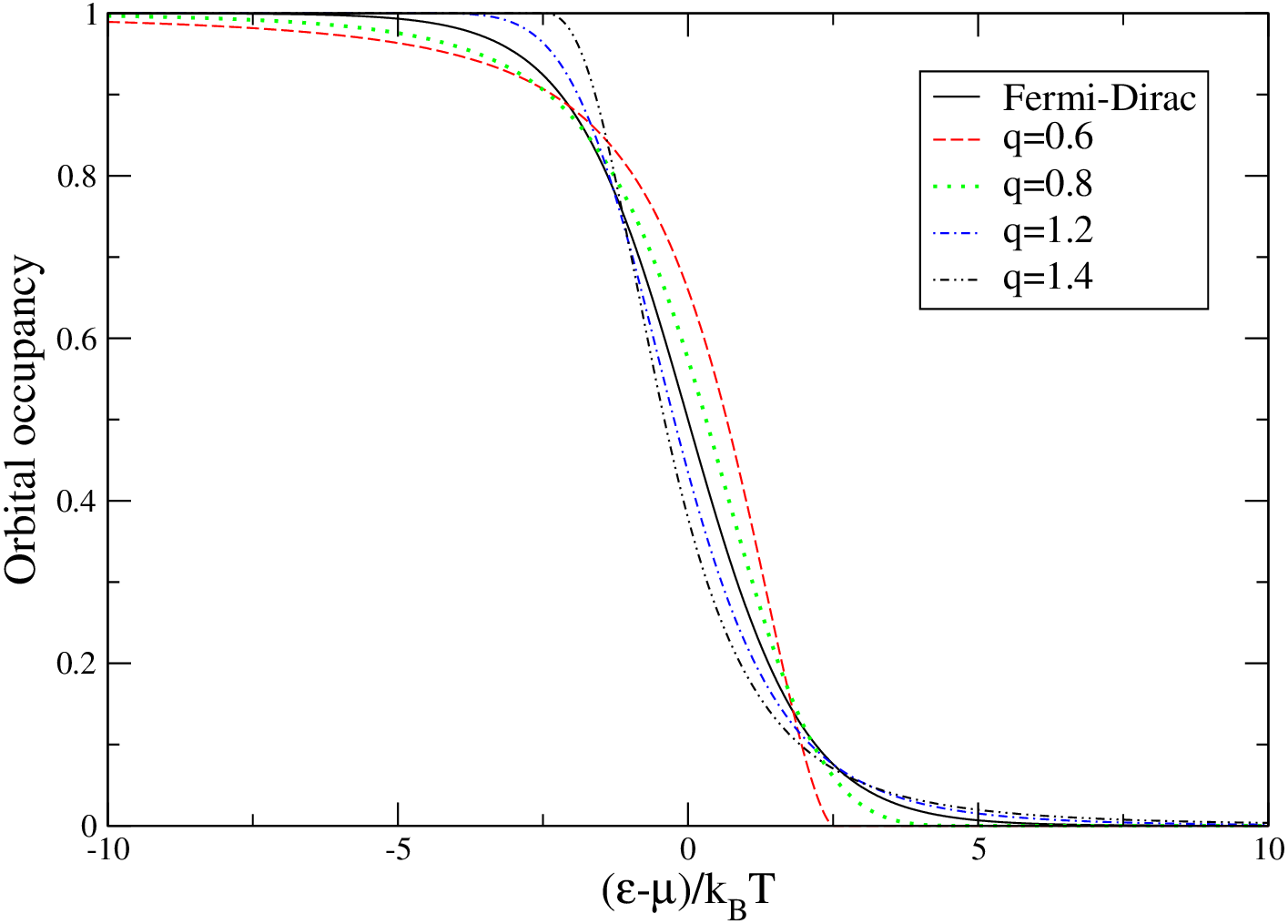}
\vspace{0.5cm}
\caption{Comparison between Fermi-Dirac distribution and the non-extensive orbital occupancies $f_q^q$ for different values of parameter $q$.}
\vspace{0.5cm}
\label{fig1}    
\end{figure}


\section{Non-extensivity of the average-atom model and Tsallis' statistics}
\label{sec2}
In order to generalize BG statistics, C. Tsallis \cite{TSALLIS88} proposed to replace the usual extensive BG entropy 

\begin{equation}
S_G=-k_B\sum_iW_i\ln W_i,
\end{equation}

$W_i$ being the probability of the state $i$ of the system, by the entropy

\begin{equation}\label{ts1}
S_{T,q}=k_B\frac{1-\sum_iW_i^q}{q-1},
\end{equation}

where $q$ is a positive real number. For instance, turbulence in electron plasmas, the flux of solar neutrinos, self-gravitating systems, bare Coulomb systems have been successfully described by Tsallis' statistics. BG statistics also fails in the interior solar plasma \cite{DU06}. Given two independent systems in the sense of factorizability of the micro-state probabilities, Tsallis' entropy of the composite system $A+B$ verifies

\begin{equation}
S_{T,q}(A+B)=S_{T,q}(A)+S_{T,q}(B)+(1-q)\frac{S_{T,q}(A)S_{T,q}(B)}{k_B}.
\end{equation}

The quantity $\left|1-q\right|$ quantifies the lack of extensivity of $S_{T,q}$ (the system is ``over-extensive'' if $q<$ 1 and ``under-extensive'' if $q>$ 1). In this formalism, the exponential function $\exp(x)$ is replaced by 

\begin{equation}
\mbox{$\exp_q(x)$}=\left\{\begin{array}{ll}
\mbox{$\left[1+(1-q)x\right]^{\frac{1}{1-q}}$} \;\; & \mbox{if \;\; $\left[1+(1-q)x\right]>0$;} \\
0 & \mbox{otherwise,}
         \end{array}
	        \right.
\end{equation}

and the logarithmic function $\ln(x)$ by

\begin{equation}
\ln_q(x)=\frac{x^{1-q}-1}{1-q}.
\end{equation}

The definition of entropy\footnote{Other $q-$type distributions exist (see for instance Refs. \cite{NADARAJAH07,MASI05}).} from Eq. (\ref{ts1}) warrants that standard BG statistics is recovered when $q\rightarrow$ 1. 

\begin{figure}
\vspace{0.5cm}
\includegraphics[width=12cm]{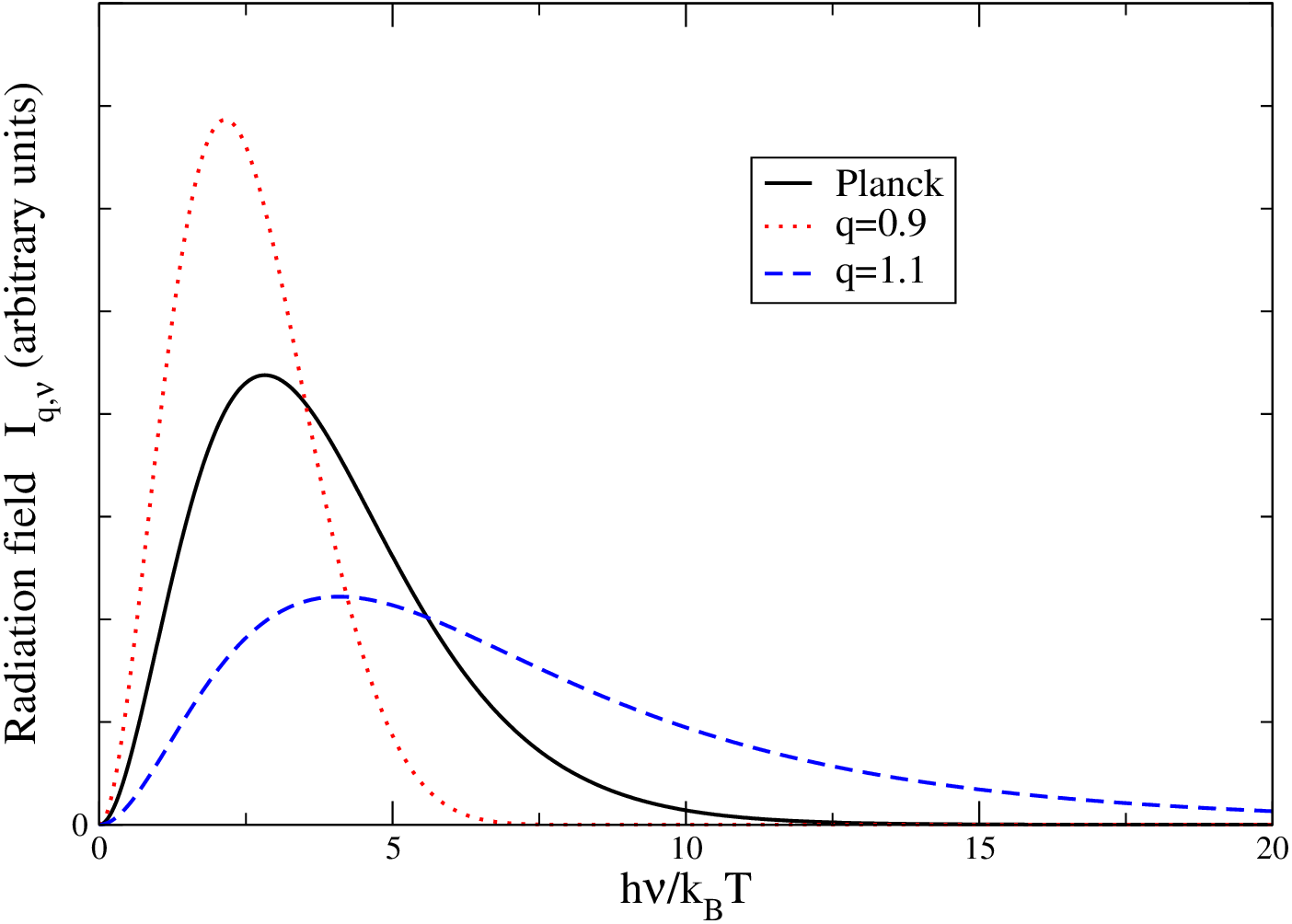}
\vspace{0.5cm}
\caption{Comparaison between Planck distribution and the non-extensive radiation distributions $I_{q,\nu}$ for different values of parameter $q$.}
\vspace{0.5cm}
\label{fig2}
\end{figure}


\section{\label{sec3}Determination of the electronic structure}

\subsection{\label{subsec31}Minimization of the non-extensive free energy}

In a non-extensive context, the number of bound electrons is given by

\begin{equation}
N_{q,b}=\sum_kg_k\left[f_{q,k}\right]^q,
\end{equation}

where $f_{q,k}$ is a factor characteristic of orbital $k$, whose degeneracy and occupancy are respectively $g_k$ and $[f_{q,k}]^q$. The internal energy reads

\begin{equation}
E_{q,b}=\sum_kg_k\left[f_{q,k}\right]^q\epsilon_{q,k},
\end{equation}

where $\epsilon_{q,k}$ is the energy of orbital $k$, and the entropy is given by

\begin{equation}\label{entq}
S_{q,b}=\frac{k_B}{q-1}\sum_kg_k\left\{1-\left[1-f_{q,k}\right]^q-\left[f_{q,k}\right]^q\right\}.
\end{equation}

It is worth mentioning that expression (\ref{entq}) comes from the fractal representation of a non-equilibrium electron gas \cite{BUY93}, which states are described by a set of orbital occupancy numbers. The total free energy of the plasma can be written

\begin{eqnarray}\label{fe}
F_q&=&F_{q,b}+F_{q,f}+F_{q,int}+F_{q,R}\nonumber\\
& &-\mu_q\left(\sum_ig_i\left[f_{q,i}\right]^q-Z+Z^*_q\right),
\end{eqnarray}

where $F_{q,b}=E_{q,b}-TS_{q,b}$ is the free energy of the bound electrons and $F_{q,R}$ the free energy of the radiation field $I_{q,\nu}$. One has

\begin{equation}
F_{q,R}=\int h\nu\Big\{\frac{c^2I_{q,\nu}}{2h\nu^3}\Big\}^q d\Omega d\nu-T\int s_{q,\nu}~d\Omega d\nu,
\end{equation}

where $h$ is Planck constant and $c$ the speed of light The entropy density of the radiation field reads

\begin{equation}
s_{q,\nu}=\frac{k_B}{q-1}\frac{2\nu^2}{c^3}\Big\{\Big[1+\frac{c^2I_{q,\nu}}{2h
\nu^3}\Big]^q-\Big[\frac{c^2I_{q,\nu}}{2h\nu^3}\Big]^q-1\Big\}.
\end{equation}

The term $F_{q,d}$ represents the free-energy of the delocalized electrons and $F_{q,int}$ the free energy accounting for interactions between electrons. The quantity $Z^*_q$ is the average ionization and $\mu_q$ is the chemical potential, determined by the preservation of the total number of electrons. Minimization of total free energy $F_q$ given by equation (\ref{fe}) with respect to $f_{q,k}$ gives \cite{BUY93,BUY95}

\begin{equation}\label{pop}
f_{q,k}=\frac{1}{\left\{1+\frac{(q-1)}{k_BT}[\epsilon_{q,k}-\mu_q]\right\}^{\frac{1}{q-1}}+1}.
\end{equation}

For $q>1$, the distribution $f_{q,k}$ gives higher values than Fermi-Dirac distribution, and lower values for $q<1$. Figure \ref{fig1} represents orbital occupancy $[f_{q,k}]^q$ for different values of $q$ compared to usual Fermi-Dirac factor $f_k$. The bound-electron density reads:

\begin{equation}\label{denb}
n_{q,b}(r)=\sum_{k}g_k\left[f_{q,k}\right]^q\left|\psi_{q,k}\right|^2.
\end{equation}

The minimization of $F_q$ with respect to the delocalized-electron density leads to

\begin{equation}\label{tfne}
n_{q,d}(r)=\frac{1}{\pi^2\hbar^3}\int_{X_q(r)}^{\infty}\frac{p^2dp}{\left\{1+\frac{(q-1)}{k_BT}\left[\frac{p^2}{2m}-\alpha_q(r)\right]\right\}^{\frac{1}{q-1}}+1},
\end{equation}

where  $V_q$ is the new electrostatic potential, 

\begin{equation}
X_q(r)=\sqrt{2m\left[\alpha_q(r)-\mu_q\right]}
\end{equation}

and 

\begin{equation}
\alpha_q(r)=\mu_q-V_q(r). 
\end{equation}

Martinenko and Shivamoggi \cite{MARTINENKO04} have shown that non-extensive effects on Thomas-Fermi (TF) model can reduce the binding energy, correcting in that way a weakness of TF model, due to the divergence $r^{-3/2}$ of TF radial density close to the nucleus, causing a breakdown in the local density approximation and a non-physical enhancement of the binding energy (boundary effect). However, their expression of the Fermi distribution differs from ours: the power is $q/(q-1)$ instead of $1/(q-1)$ in our expression. Moreover, here we have uncomplete Fermi integrals. Equation (\ref{tfne}) can be written

\begin{equation}\label{tfne2}
n_{q,f}(r)=\frac{\sqrt{2}}{\pi^2\hbar^3}(mk_BT)^{3/2}F_{1/2}(\frac{\alpha_q(r)-\mu_q}{k_BT},\frac{\alpha_q(r)}{k_BT},q),
\end{equation}

where

\begin{equation}
F_n(a,x,q)=\int_{a}^{\infty}\frac{t^n}{1+\left[1+(q-1)(t-x)\right]^{\frac{1}{q-1}}}dt,
\end{equation}

and the particular integral $F_{1/2}(a,x,q)$ is convergent only for $q<5/3$. At low temperatures ($x\gg 1$), one has

\begin{equation}
F_n(x\gg 1)\approx\frac{x^{n+1}}{n+1}\int_{-\infty}^{\infty}
\frac{\left(1+\frac{t}{x}\right)^n\left[1+(q-1)t\right]^{\frac{1}{q-1}}}{1+\left[1+(q-1)t\right]^{\frac{1}{q-1}}}dt,
\end{equation}

which can be written

\begin{equation}
F_n(x\gg 1)\approx\frac{x^{n+1}}{n+1}~\zeta_0(q)+x^n~\zeta_1(q)+\frac{nx^{n-1}}{2}~\zeta_2(q)+\cdots
\end{equation}

where $\zeta_j$ is defined by

\begin{equation}
\zeta_j(q)=\int_{-\infty}^{\infty}\frac{t^j\left[1+(q-1)t\right]^{\frac{1}{q-1}}}
{1+\left[1+(q-1)t\right]^{\frac{1}{q-1}}}dt.
\end{equation}

The minimization of the total free energy defined by Eq. (\ref{fe}) with respect to the radiation field $I_{q,\nu}$ gives

\begin{equation}\label{qpla}
I_{q,\nu}=\frac{2h\nu^3}{c^2}\frac{1}{\left[1
+\frac{(q-1)}{k_BT}h\nu\right]^{\frac{1}{q-1}}-1},
\end{equation}

the non-extensive radiation field, which does not follow Planck law. Figure \ref{fig2} represents $I_{q,\nu}$ for different values of $q$, compared to usual Planck distribution. For $q$ lower (greater) than 1, the maximum is higher (lower) than for Planck distribution, and slightly shifted to lower (higher) energy. In other words, the energy $h\nu_0$ at which the absorption of photons is maximum is larger than 2.822 $k_BT$ if $q>$ 1 and lower than 2.822 $k_BT$ if $q<$ 1.

\vspace{1cm}

\begin{figure}
\vspace{0.5cm}
\includegraphics[width=12cm]{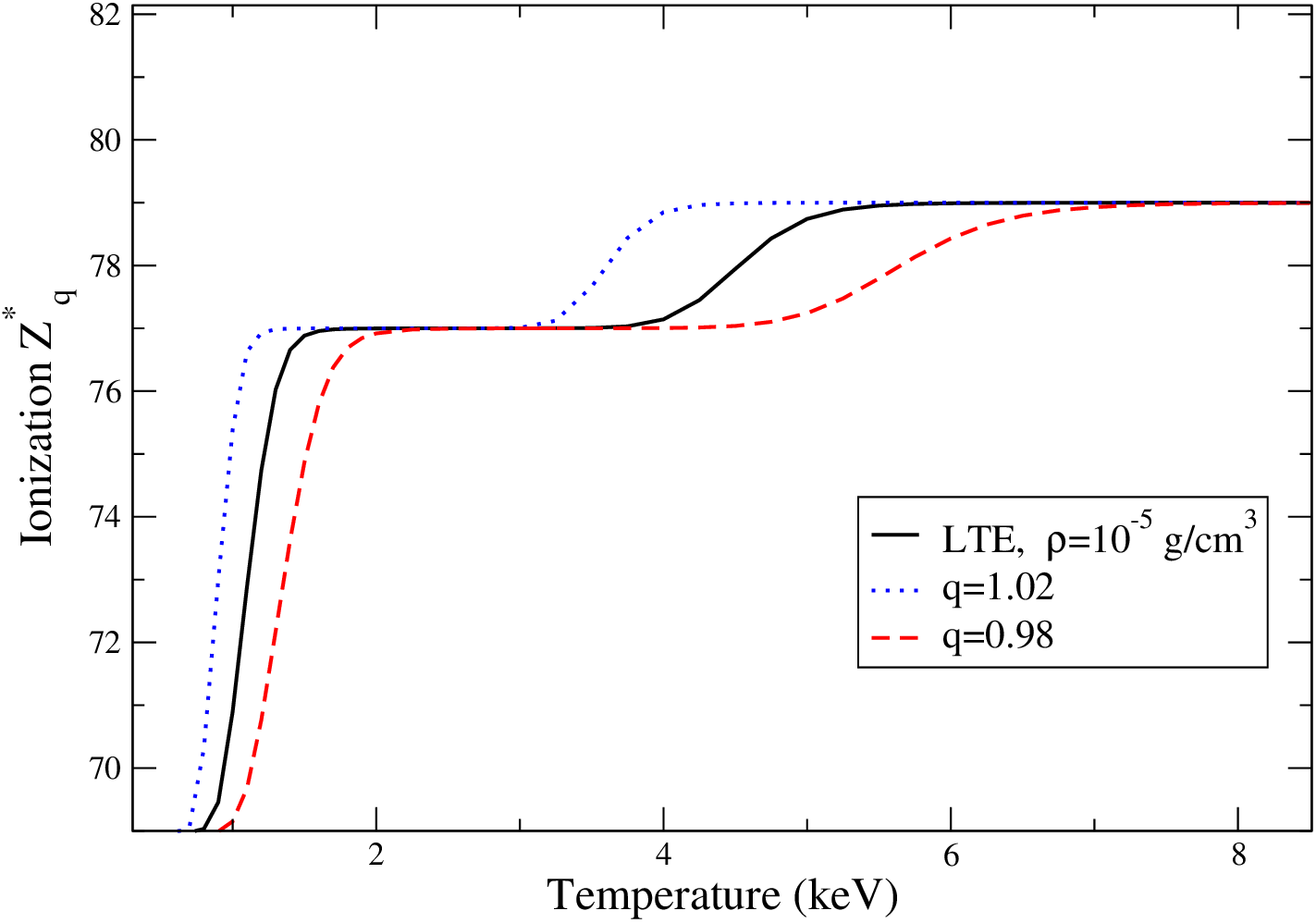}
\vspace{0.5cm}
\caption{Ionization versus temperature for a gold plasma at $\rho=10^{-5}$
g/cm$^3$ from an LTE calculation and from non-extensive calculations with
$q$=0.98 and 1.05.}
\vspace{0.5cm}
\label{fig3}
\end{figure}

\begin{figure}
\vspace{0.5cm}
\includegraphics[width=12cm]{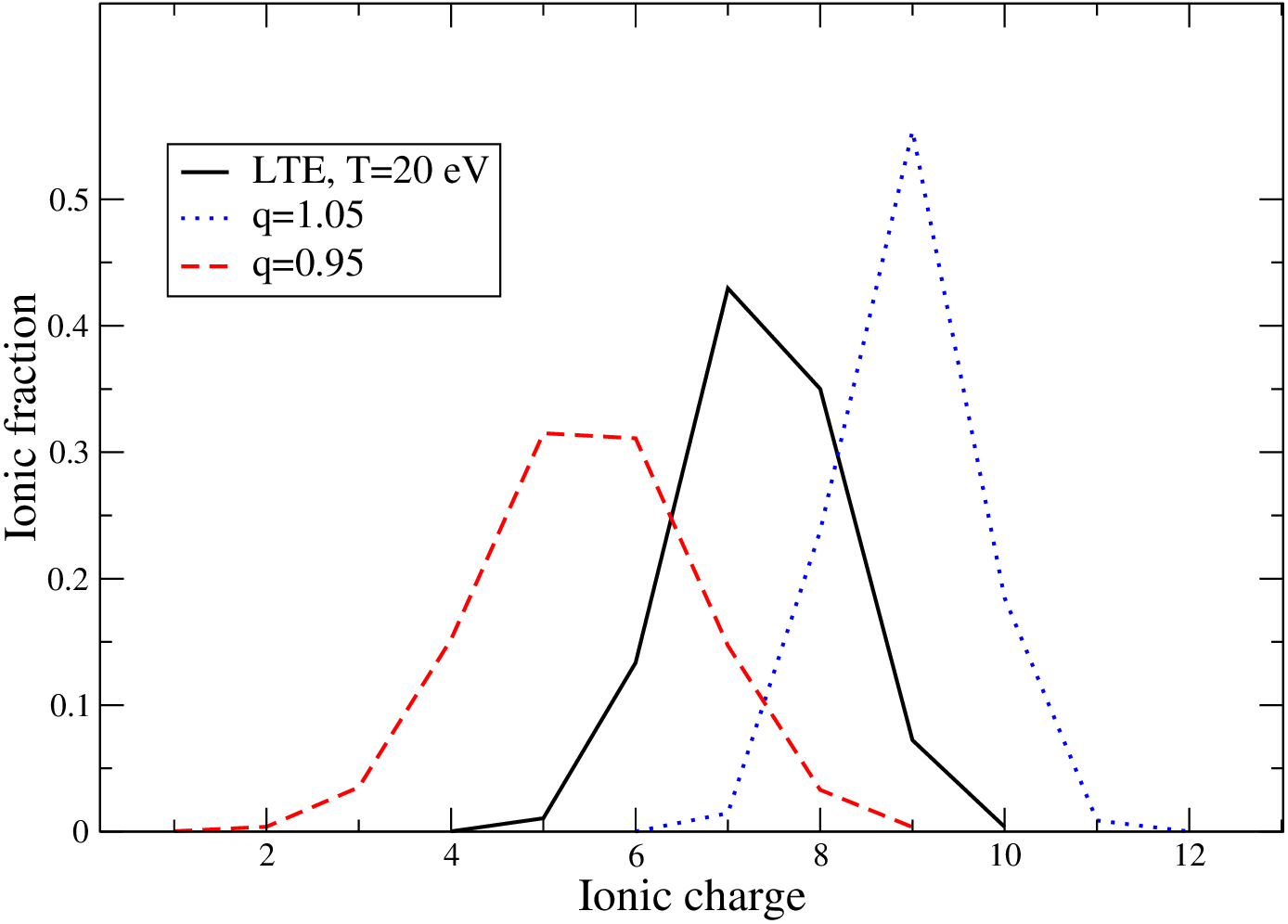}
\vspace{0.5cm}
\caption{Ionic distributions of an iron plasma at $\rho$=10$^{-3}$ g/cm$^3$ and
$T$=20 eV from an LTE calculation and from non-extensive calculations with
$q$=0.95 and 1.05.}
\vspace{0.5cm}
\label{fig4}
\end{figure}

\begin{figure}
\vspace{0.5cm}
\includegraphics[width=12cm]{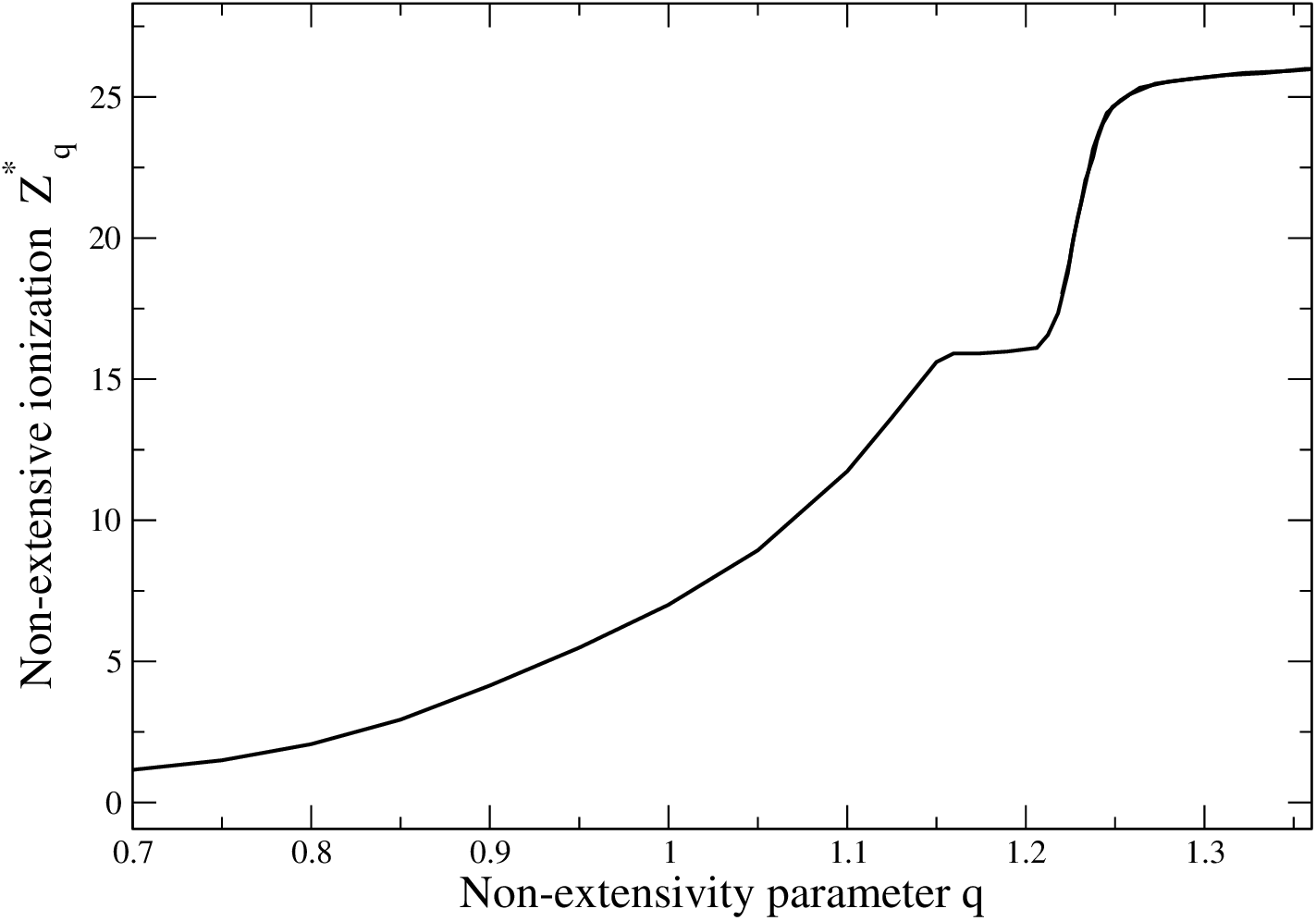}
\vspace{0.5cm}
\caption{Ionization for different values of $q$ for an iron plasma at
$\rho$=10$^{-3}$ g/cm$^3$ and $T$=20 eV.}
\vspace{0.5cm}
\label{fig5}
\end{figure}

\vspace{1cm}

\subsection{\label{subsec32}Determination of the self-consistent potential and electro-neutrality}

The bound states are obtained solving Schr\"odinger equation and relativistic effects are taken into account in the Pauli approximation \cite{BLENSKI95}. The new electrostatic potential $V_q$ can be written:

\begin{equation}
V_q(r)=V_{q,c}(r)+V_{q,xc}(r),
\end{equation}

where $V_{q,xc}$ represents the finite-temperature exchange-correlation potential evaluated following \cite{ICHIMARU87} in the local density approximation (at electron density $n_q(r)=n_{q,b}(r)+n_{q,f}(r)$). The coulombic part $V_{q,c}$ is obtained from Poisson equation

\begin{equation}
\Delta V_{q,c}(r)=-\frac{e^2}{\epsilon_0}\left[n_{q,b}(r)+n_{q,f}(r)\right],
\end{equation}

where $n_{q,b}$ and $n_{q,f}$ are given respectively by Eqs. (\ref{denb}) and (\ref{tfne2}). The quantity $\epsilon_0$ represents the permittivity of vacuum and one has:

\begin{equation}
V_{q,c}(r)\underbrace{\approx}_{r\rightarrow 0}-\frac{Ze^2}{4\pi\epsilon_0r}.
\end{equation}

The chemical potential $\mu_q$ is determined from the neutrality of the average-ion spherical cell:

\begin{equation}\label{cons}
N_{q,b}=Z-Z^*_q,
\end{equation}

where the average charge $Z^*_q$ is evaluated by integration of the delocalized-electron density defined in Eq. (\ref{tfne2}). The same process is repeated until convergence is reached, {\emph{i.e.} until the potential variation becomes as small as required. It is worth mentioning that the use of orbital occupancy factors $[f_{q,k}]^q$ can also be helpful from a numerical point of view \cite{WILSON07}. Indeed, average-atom models can suffer from convergence problems at low temperature, due to the fact that the Fermi-Dirac distribution tends to a Heaviside function. Such difficulties might be avoided performing the self-consistent calculation with $[f_{q,k}]^q$ distribution, and taking the limit $q\rightarrow 1$.


\section{\label{sec4}Orbital energies, populations and average ionization}

Tables \ref{tab1} and \ref{tab2} display respectively the energy and population of several orbitals for an iron plasma at $\rho$=10$^{-3}$ g/cm$^3$ and $T$=20 eV from an LTE calculation and from non-extensive calculations with $q$=0.95 and 1.05. When $q<$ 1 (respectively $q>$ 1), energies and populations are higher (respectively lower) than in the LTE case. Figure \ref{fig3} represents the average ionization versus temperature for a gold plasma at $\rho$=10$^{-5}$ g/cm$^3$ in LTE and in the non-extensive formalism for $q$=0.98 and 1.02. We can see that the non-extensive ionization is lower than the extensive one for $q<$ 1 and larger for $q>$ 1. Figure \ref{fig4} represents the ionic distributions in an iron plasma at $\rho$=10$^{-3}$ g/cm$^3$ and $T$=20 eV in the LTE case and calculated with the $q-$version of the code for $q$=1.05 and $q$=0.95. The non-extensive distributions are shifted to lower charge states (if $q<$ 1) and higher (if $q>$ 1) and their asymetry can be different from the LTE one: the ionic distribution is wider for $q<$ 1. The difference increases with the value of $q$. Figure \ref{fig5} shows non-extensive ionization $Z^*_q$ versus non-extensivity parameter $q$. One can notice two steps: the first one corresponds to $Z^*_q$=16, \emph{i.e.} 10 remaining bound electrons, corresponding to full K and L shells. The second step corresponds to the asymptote of full ionization $Z^*_q\rightarrow Z$=26.

\begin{table}
\begin{center}
\begin{tabular}{cccc} \hline
Orbital & Energy  & Energy & Energy \\ 
        & at LTE & for $q$=0.95 & $q$=1.05 \\ \hline
3s & -238.90 & -205.71 & -289.51 \\ 
$3p_{1/2}$ & -204.89 & -171.97 & -255.77 \\ 
$3d_{3/2}$ & -146.66 & -114.01 & -198.36 \\ 
$4d_{5/2}$ & -60.95 & -43.54 & -88.98 \\ 
$6f_{5/2}$ & -16.60 & -10.07 & -27.75 \\ 
$8h_{11/2}$ & -5.99 & -2.99 & -11.16 \\ \hline
\end{tabular}
\end{center}
\caption{Energies (eV) of a few selected orbitals for different approaches in
the case of an
iron plasma at $\rho$=10$^{-3}$ g/cm$^3$ and $T$=20 eV: LTE case and non-extensive
cases with
$q$=0.95 and $q$=1.05.}\label{tab1}
\end{table}

\begin{table}(1995)
\begin{center}
\begin{tabular}{cccc} \hline
Orbital & Population & Population & Population \\
        & at LTE & for $q$=0.95 & for $q$=1.05 \\ \hline
3s & 1.92 & 1.93 & 1.84 \\ 
3p$_{1/2}$ & 1.62 & 1.69 & 1.38 \\ 
3d$_{3/2}$ & 0.77 & 0.92 & 0.45 \\ 
4d$_{5/2}$ & 0.02 & 0.05 & 0.003 \\ 
6f$_{5/2}$ & 0.002 & 0.0099 & 0.0001 \\ 
8h$_{11/2}$ & 0.002 & 0.014 & 0.0001 \\ \hline
\end{tabular}
\end{center}
\caption{Population (number of electrons) of a few selected orbitals for different approaches in the
case of an iron
plasma at $\rho$=10$^{-3}$ g/cm$^3$ and $T$=20 eV: LTE case and non-extensive cases
with $q$=0.95
and $q$=1.05.}\label{tab2}
\end{table}


\section{\label{sec5}Conclusion}

In order to investigate the impact of non-extensivity on atomic physics in hot plasmas, a self-consistent atomic-structure model was presented, resulting from the minimization of a non-extensive free-energy and including the radiation field. All electrons (bound and delocalized) are described in the framework of the non-extensive formalism. Delocalized electrons are taken into account using non-extensive semi-classical Thomas-Fermi approximation, which leads to a reduction of the binding energy, correcting in that way a weakness of standard Thomas-Fermi model. Standard thermodynamics is recovered when $q\rightarrow 1$ and the formalism presented here can be extended to other approaches of hot-plasma atomic structure. For instance, another possibility to account for the electron states consists in considering not only the mean configuration of the plasma, but the real configurations (with integer electron populations for the orbitals). Because the number of configurations is huge, they can be gathered into superconfigurations \cite{PAIN02,PAIN03,PAIN06a,PAIN06b,PAIN07b} and the self-consistent calculation of electronic structure presented in this paper can be performed for each superconfiguration. In view of the growing interest for Tsallis' non-extensive statistics, we hope that the present study will give rise to new ideas for the simulation of plasma atomic physics.

\end{document}